%% file: manuscript.tex
\documentclass[preprints,article,accept,oneauthor,pdftex]{Definitions/mdpi} 

\usepackage{physics}

\usepackage{graphicx}
\usepackage{color}
\usepackage{xcolor}
\usepackage{tikz,pgfplots}
\usetikzlibrary{positioning,fit,calc}
\usetikzlibrary{shapes.arrows,arrows, decorations.markings}
\usetikzlibrary{plotmarks}
\pgfplotsset{compat=newest}

\usepackage{caption}
\usepackage{subcaption}
\usepackage{multirow}

\newcommand{\code}[1]{\emph{#1}}
\newcommand{\comment}[1]{}

\definecolor{color1}{HTML}{8DD3C7}
\definecolor{color2}{HTML}{FFFFB3}
\definecolor{color3}{HTML}{BEBADA}
\definecolor{color4}{HTML}{FB8072}
\definecolor{color5}{HTML}{80B1D3}
\definecolor{color6}{HTML}{FDB462}
\definecolor{color7}{HTML}{B3DE69}
\definecolor{color8}{HTML}{FCCDE5}
\definecolor{color9}{HTML}{D9D9D9}
\definecolor{color10}{HTML}{BC80BD}
\definecolor{color11}{HTML}{CCEBC5}
\definecolor{color12}{HTML}{FFED6F}
\definecolor{color21}{HTML}{A6CEE3}
\definecolor{color22}{HTML}{1F78B4}
\definecolor{color23}{HTML}{B2DF8A}
\definecolor{color24}{HTML}{33A02C}
\definecolor{color25}{HTML}{FB9A99}
\definecolor{color26}{HTML}{E31A1C}
\definecolor{color27}{HTML}{FDBF6F}
\definecolor{color28}{HTML}{FF7F00}
\definecolor{color29}{HTML}{CAB2D6}
\definecolor{color210}{HTML}{6A3D9A}
\definecolor{color211}{HTML}{FFFF99}
\definecolor{color212}{HTML}{B15928}
\definecolor{gray50}{rgb}{0.5,0.5,0.5}
\definecolor{gray75}{rgb}{0.75,0.75,0.75}

\usepackage[linesnumbered,ruled,vlined]{algorithm2e}

\firstpage{1} 
\pubvolume{1}
\issuenum{1}
\articlenumber{0}
\pubyear{2021}
\copyrightyear{2021}
\datereceived{} 
\dateaccepted{} 
\datepublished{} 
\hreflink{https://doi.org/} 
\pdfoutput=1

\Title{Data-Oriented Language Implementation of Lattice-Boltzmann Method for Dense and Sparse Geometries}

\TitleCitation{Data-Oriented Language Implementation of Lattice-Boltzmann Method for Dense and Sparse Geometries}

\Author{Tadeusz Tomczak \orcidA{}}

\AuthorNames{Tadeusz Tomczak}

\AuthorCitation{Tomczak, T.}

\address{%
Department of Computer Engineering, Wrocław University of Science and Technology, 50-370 Wrocław, Poland; tadeusz.tomczak@pwr.edu.pl\\}

\abstract{
The performance of lattice-Boltzmann solver implementations usually depends mainly on memory access patterns.
Achieving high performance requires then complex code which handles careful data placement and ordering of memory transactions.
In this work, we analyse the performance of an implementation based on a new approach called the data-oriented language, which allows the combining of complex memory access patterns with simple source code.
As a use case, we present and provide the source code of a solver for D2Q9 lattice and show its performance on GTX Titan Xp GPU for dense and sparse geometries up to $4096^2$ nodes.
The obtained results are promising, around 1000 lines of code allowed us to achieve performance in the range of 0.6 to 0.7 of maximum theoretical memory bandwidth (over 2.5 and 5.0 GLUPS for double and single precision, respectively) for meshes of size above $1024^2$ nodes, which is close to the current state-of-the-art.
However, we also observed relatively high and sometimes difficult to predict overheads, especially for sparse data structures.
The additional issue was also a rather long compilation, which extended the time of short simulations, and a lack of access to low-level optimisation mechanisms.
}


\begin{document}

\section{Introduction}

Current high-performance computers use some form of parallel processing on many levels: beginning at instruction-level parallelism (ILP) and single instruction multiple data (SIMD) support, through the use of dynamic random access memories (DRAM), which transfer data in blocks containing several dozen bytes, up to multi/many-core chips and clusters of machines connected with a fast network.
Thus, to effectively use the available hardware, the processed data should be carefully arranged in a way that allows usage of all available hardware with minimal losses.
For example, DRAM block transactions connected with SIMD processing are tuned to large data sets containing elements processed in the same way.
When neighbouring elements require different operations, the hardware usually is significantly underutilized.
These limitations cause that many computational problems, for example in the physic simulations area, require not only sophisticated algorithms but also non-trivial data layouts in memory to achieve high performance.
Typical examples of such problems are simulations on sparse geometries, i.e. geometries for which computations must be done only for a small part of area/volume.

The lattice-Boltzmann method (LBM) is a computational fluid dynamics (CFD) algorithm based on cellular automata idea, where automaton cells correspond to points (called \emph{nodes}) of a uniformly discretized domain of computations.
One of the main advantages of LBM is its inherent parallelism, thus many high-performance LBM implementations 
are known.
For dense geometries, the implementation may be relatively simple \cite{Tolke:Imp2008, Tolke:Tera2008} and allow to achieve high hardware utilisation (up to almost 80\% of peak theoretical memory bandwidth) 
\cite{Mawson:Mem2014, Januszewski:Sail2014}.
However, when the significant part of geometry is solid and many nodes of a discretized domain do not take part in computations, then the more complex implementation techniques have to be used to avoid memory, bandwidth, and computational power waste.

LBM implementations for sparse geometries are based on two main approaches: indirect addressing, where each node contains additional information about localisation of neighbouring nodes, and spatial discretisation, 
where the information about geometry sparsity is stored for fragments containing a number of neighbouring nodes from the domain.
Two main indirect addressing approaches are the \emph{connectivity matrix} \cite{Schulz:Par2002}, used in MUPHY \cite{Bernaschi:MUP2009} and ILBDC \cite{Zeiser:Ben2009} solvers, and the \emph{fluid index array} \cite{Martys:Mul2002, Nita:GPU2013, Januszewski:Sail2014}.
Spatial discretisation is used in HemeLB \cite{Mazzeo:Hem2008}, Palabos \cite{Parmigiani:Por2011}, OpenLB \cite{Krause:Ope2021}, Musubi \cite{Hasert:Com2014} and WaLBerla \cite{Feichtinger:WaL2011} platforms.
A more detailed review of LBM implementations for sparse geometries can be found in \cite{Tomczak:Lat2018}.

Although there are known many techniques that allow increasing the utilisation of hardware resources for computations on sparse geometries, implementation of these techniques requires a significant amount of work, especially for different machines.
Even the rather simple implementation of one-level spatial discretisation from \cite{Tomczak:Spa2018, Tomczak:Ane2019} requires more than ten thousand lines of heavily templated C++ code.
In this context, an interesting approach to reduce the amount of work required to port the code to different platforms was presented in \cite{Januszewski:Sail2014}, where the final code for target machines was generated from templates written in the Python Mako library.

Recently, the data-oriented parallel programming language Taichi appeared \cite{Hu2019:Tai}, which simplifies the development of high-performance codes for computations on both sparse and dense data structures.
It allows not only to generate the final code for different target machines, but the more important feature is decoupling information about data structures from computational kernels.
By providing \emph{structural nodes}, which can be used to build complex, hierarchical data structures, and at the same time offer a simple interface simulating access through \verb|[]| operator like for dense data structures, the Taichi language allows to design only simple, basic codes describing computations only.

In this work, we present an implementation of the lattice-Boltzmann method in Taichi language for both dense and sparse geometries and investigate its performance on a massively parallel graphic processing unit (GPU).
To our knowledge, currently, there are no published studies on Taichi-based LBM solvers.
A simple code of one existing attempt is available \cite{LBMTaichi},
but it was designed to keep the code simple, thus its performance is low.
The implementation presented in this work is loosely based on this simple version, but we significantly redesigned the code to improve performance and handle a wider set of boundary conditions.
Our implementation is created mainly for performance analysis, although we did some elementary correctness tests.
We also provide the source code\footnote{Available at \url{https://github.com/tadeusz-tomczak/tilb}} which, as we believe, can be a good starting point for building high-performance simulations of more complex physical phenomena.

\section{Materials and Methods}

\subsection{Taichi language}

The Taichi programming language \cite{Hu2019:Tai} is an actively developed open-source \cite{taichi}
just-in-time compiler that translates Python-like source code to the binary code for different hardware platforms (various CPUs, CUDA, AMDGPU and others).
The language uses the standard Python syntax extended with decorators marking some functions as \emph{kernels}.
During compilation, the kernels are transformed to optimised binary codes, which are then called from the surrounding Python source.
Such a solution allows to easily bind efficient, application-specific kernels with a wide variety of available libraries and tools, especially that Taichi provides a basic GUI system and simple built-in interfaces to NumPy and PyTorch libraries.

Kernel compiler applies few levels of optimisations, including simple template instantiation, loop unrolling and vectorization, constant folding, and others.
Inside kernels, the highest level loops can be automatically parallelised provided that operations are done on Taichi \emph{fields}.

Fields are the multidimensional data structures that provide an array-like data access interface via the \verb|[]| operator.
Internally fields use a hierarchy of \emph{structural nodes} (SNodes) which define dimensions, size and data arrangement in memory.
Information from SNodes is used, both during kernel compilation and runtime, to optimise data access to field elements and to distribute workload onto available threads/processors.
Compilation-time optimisations allow decreasing overheads caused by traversing nested SNodes but require that field definition must be known at the moment of kernel compilation and enforce separate kernel compilation for different data structures.

The types of SNode allow using different data layouts.
The \code{dense} layout is a simple, multidimensional array.
The \code{bitmasked} layout allows adding information whether given elements contain valid data or not.
This layout does not reduce memory usage because all bitmasked elements must be still placed in memory.
However, computational kernels are called only for data elements masked as valid, thus the \code{bitmasked} layout can reduce the number of operations for sparse data.
Reduction of memory usage for sparse data is possible with the \code{pointer} data layout that can mark pointers to non-existent data as invalid and thus does not require memory allocation in this case.

The data layout in Taichi can be defined as a hierarchy of different SNodes.
For example, \code{ti.root.pointer (ti.i, 16).bitmasked (ti.i, 8).dense (ti.i, 4)} defines 16 pointers, each pointing to 8 bitmasked dense blocks of data where each dense block contains 4 data elements (not shown).
After definition, this structure can be accessed using simple \code{[i]} operator, where $\mathrm{i} \in \lbrace 0 \ldots 511 \rbrace$.
The calculations of memory addresses, traversing and checking values of pointers and bitmasks, and launching the appropriate number of computational kernels are internally handled by Taichi.
Additionally, some optimisations are applied to reduce the overheads caused by additional memory accesses required to traverse multi-level, hierarchical data structures.

\subsection{Lattice-Boltzmann method}

The lattice-Boltzmann method (LBM) is a numerical approach to solve the Navier-Stokes equations, which describe the motion of fluids.
The detailed LBM description with the theoretical background is available in many books \cite{Kruger:The2017, Guo:Adv2013, Mohamad:Lat2019}, thus in this work, we only show a minimal introduction from the implementation point of view.

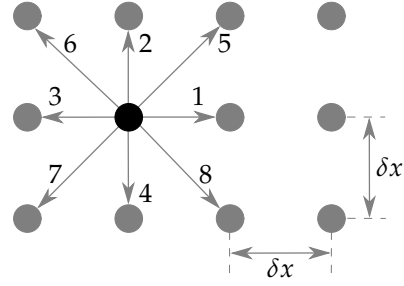
\begin{figure}
\begin{center}
\input{img/la_d2q9.tex}
\caption{D2Q9 lattice arrangement and indices of lattice links (index 0 deontes the node itself). Circles denote lattice nodes.}
\label{fig:lat_d2q9}
\end{center}
\end{figure}

In LBM, the domain is discretized into a uniform, usually Cartesian, mesh containing \emph{nodes} distant by the \emph{lattice spacing} $\delta x$ along all axes.
During computations, nodes communicate with some neighbours - the choice of neighbours depends on \emph{lattice arrangement} which defines the dimension of the problem and communication pattern between neighbouring nodes (\emph{lattice linkage}).
The lattice arrangement is usually described using D$d$Q$q$ notation, where $d$ is the dimension, and $q$ is the linkage.
For example, the D2Q5 arrangement defines a 2D lattice where nodes communicate only with neighbours placed along the axes (left, right, top, bottom), whereas D2Q9 takes into account also the nodes placed at diagonal corners (see Figure \ref{fig:lat_d2q9}).

A single iteration of simulation advances simulation time by the time step $\delta t$ and includes one-time communication between all nodes and additional computations.
To simplify equations, it is usually assumed that $\delta x = 1$ and $\delta t = 1$.
In such case, the initial \emph{characteristic velocity} $U$ and the fluid viscosity $\nu$ have to be set up to keep the required Reynolds number
\begin{equation}
	\mathrm{Re} = \frac {U \cdot L}{\nu}
	,
	\label{eqn_Re}
\end{equation}
where $L$ denote the \emph{characteristic length} which is dependent on the selected size in simulated geometry.

Each node contains a set of \emph{particle distribution functions} (PDF) $f_i (\vb{x},t)$, where $\vb{x}$ denotes node position, $t$ denotes time, and $i$ denotes the index of function corresponding to lattice linkage.
The PDF numbering can be chosen in different ways, in this work we use one of the most often presented in Figure \ref{fig:lat_d2q9}.
The macroscopic fluid density $\rho$ and velocity $\vb{v}$ are related with PDFs according to equations
\begin{equation}
\rho (\vb{x},t) = \sum_i f_i (\vb{x},t)  \quad \mathrm{and} \quad  \vb{v} (x,t) = \frac{1}{\rho (\vb{x},t)} \frac{\delta x}{\delta t} \sum_i \vb{v_i} f_i (\vb{x}, t)
,
	\label{eqn_rho_u}
\end{equation}
where $\vb{v_i}$ are called \emph{lattice (microscopic) velocities} and are equal to vectors from current to neighbour node.
For D2Q9 lattice shown in Figure \ref{fig:lat_d2q9} the selected values of $\vb{v_i}$ are the following: $\vb{v_0} = [0,0], \vb{v_1} = [1, 0], \vb{v_4} = [0, -1], \vb{v_6} = [-1, 1]$ etc.

The LBM operations are described by equation
\begin{equation}
	f_i (\vb{x} + \vb{v_i} \delta t, t + \delta t) = f_i (\vb{x}, t) + \Omega_i,
	\label{eqn_LBM}
\end{equation}
where $\Omega_i$ is named the \emph{collision operator}.
Simple LBM implementations are then often realised as two alternating steps: \emph{collision} computing new PDF values according to the right-hand side of Eqn. (\ref{eqn_LBM}), and \emph{streaming} responsible for transferring the values computed during the collision step into the places defined by the left-hand side of Eqn. (\ref{eqn_LBM}).

One of the simple, yet widely used, collision operators is the Bhatnagar, Gross and Krook (BGK) operator \cite{AMo:Bhatnagar1954} defined as
\begin{equation}
	\Omega_i^{BGK} = - \omega \Big( f_i (\vb{x}, t) - f_i^{eq} (\vb{x}, t) \Big)
	,
	\label{eqn_OBGK}
\end{equation}
where $\omega$ is called the \emph{collision frequency} and $f_i^{eq}$ is the \emph{local equilibrium distribution function}.
Assuming both $\delta{x}$ and $\delta t$ equal to one, the collision frequency results from the fluid viscosity $\nu$ as
\begin{equation}
	\omega = \frac{1}{3  \nu + \frac{1}{2}}
	.
\end{equation}
The local equilibrium distribution function is defined as
\begin{equation}
	 f_i^{eq} (\vb{x}, t) = w_i \rho \Big( 1 + 3 \vb{v_i} \cdot \vb{v} + \frac{9}{2} (\vb{v_i} \cdot \vb{v})^2 - \frac{3}{2} \vb{v} \cdot \vb{v} \Big)
	 ,
	 \label{eqn_feq}
\end{equation}
where constants $w_i$ depend on lattice arrangement and for D2Q9 are $w_0 = \frac{4}{9}, w_{1,2,3,4} = \frac{1}{9}, w_{5,6,7,8} = \frac{1}{36}$.

\section{Implementation}

The LBM implementation can directly follow Eqn. (\ref{eqn_LBM}) but in our version, we applied some of the typical, widely known techniques developed for higher performance.
The general idea of the presented implementation is shown in Algorithm \ref{alg_LBM}.
The algorithm contains three main operations.

\begin{algorithm}[t]

	\For {all nodes}
	{
		Set initial values of $\rho, \vb{v}$

		Set $f_i^{pre} = f_i^{eq}$
	}

	\SetKwBlock{StreamAndCollide}{Kernel stream and collide}{}
	\SetKwFor{ParFor}{for}{do in parallel}{}
	\SetKw{Gather}{gather}
	\SetKw{Compute}{compute}
	\SetKw{Store}{store}
	\SetKw{Swap}{swap}

	\Repeat {end of simulation}
	{

		\ParFor{all nodes at positions $\vb{x}$}
		{
			\StreamAndCollide
			{
				\For {all directions $i$}
				{
					\Gather  $\  f'_i \  \leftarrow \  f_i^{pre} (\vb{x} - \vb{v_i})$
				}
				
				\Compute $\rho$ and $\vb{v}$ from $f'_i$ using Eqn. (\ref{eqn_rho_u})

				\For {all directions $i$}
				{
					\Compute $f^{eq}_i$ from $\rho$ and $\vb{v}$ using Eqn. (\ref{eqn_feq})

					\Compute new $f''_i$ from $f'_i, f^{eq}_i$, RHS of Eqn. (\ref{eqn_LBM}), and Eqn. (\ref{eqn_OBGK})

					\Store  $\  f^{post}_i (\vb{x}) \  \leftarrow \  f''_i$
				}
			}
		}

		\ParFor{all nodes at positions $\vb{x}$}
		{
			\Swap $\  f^{post}_i (\vb{x}) \  \rightleftarrows \  f^{pre}_i (\vb{x})$
		}

	}

	\caption{
	General idea of the presented LBM implementation. $\rho, \vb{v}, f'_i$ and $f''_i$ denote temporary values used only inside the kernel.
}
\label{alg_LBM}
\end{algorithm}

Lines 1-3 are responsible for the initialization of PDFs values $f_i$ that are set to equilibrium state $f_i^{eq}$ computed from initial values of velocity and density.
Initialization is done only once at the beginning of computations.
Then, the simulation comes down to computing values of PDFs for the successive time steps.

A single iteration of lines 5--15 corresponds to a single time step.
During the time step computations, all nodes are processed in parallel.
To avoid race conditions, we use two copies of $f_i$ functions: $f_i^{pre} = f_i (t)$ functions were computed during the previous time step and are only read, and $f_i^{post} = f_i(t + \delta t)$ functions are computed during the current time step and are only written.
There are also known parallel implementations that use a single copy of PDFs only \cite{Acc:Bailey2009} at the cost of increased code complexity.

To minimise the memory bandwidth, we implemented the \emph{fused} kernel (lines 6--13), where collision and streaming are done in one step with a single read and write of $f_i$ values.
Additionally, we use the reversed order of collision and streaming, also known as the \emph {pull} scheme \cite{Rinaldi:ALa2012}.
Direct implementation of Eqn. (\ref{eqn_LBM}) computes the collision first and then \emph{scatters} the new, computed $f_i$ values to neighbour nodes.
In the pull approach, first, the $f_i$ values from the previous time step are \emph{gathered} from neighbour nodes, then the collision is applied, and, eventually, the new $f_i$ values are stored in the current node.
This scheme keeps addresses of all writes to memory aligned what may additionally decrease memory traffic because unaligned memory writes often are more costly than unaligned reads (for example due to allocate-on-write policy).

After processing of all nodes, the values of  $f_i^{pre}$ and $f_i^{post}$ are exchanged in lines 14--15.
However, since we have not found an efficient way to exchange fields, then we use two kernels with identical computations, but one of the kernels reads  $f_i^{pre}$ and writes $f_i^{post}$ and the second kernel does the opposite.

The operations shown in Algorithm \ref{alg_LBM} do not include support for boundary conditions, which is also present in the implemented kernel.
To detect boundary nodes, we store in memory an additional field encoding each node type (fluid, solid, boundary type) along with a bitmask containing information about which neighbour nodes are present.
We also use a separate field to store values for boundary nodes with fixed conditions, e.g. constant velocity.
Supported boundary conditions are constant velocity, constant pressure, and bounce back according to \cite{Zou:Onp1997}.

In addition to the optimisations mentioned above, we also applied a few low-level optimisations: we used the structure-of-arrays instead of the array-of-structures data layout, minimised the number of memory operations and placed them in a non-divergent code to allow coalescing, and used numeric constants reducing the number of floating-point arithmetic operations.
These optimisations were applied after analysis of the generated GPU assembly.
Also, we resigned from encapsulating functionalities inside Python classes due to a small drop in performance.
Additionally, the whole code (except auxiliary functions) is placed in a single source code file what simplified the management of memory allocation and kernel generation.
The complete code contains slightly more than 1000 source lines of code, including geometry generation and storage of results.

For convenience, we also allocate memory for values of velocity $\vb{v}$ and density $\rho$, although these are not used during computations but only for data initialization, storage of results, and visualisation.
To allow run-time generation of images illustrating velocity fields, we also allocate an additional field for image memory, although it can be removed when not used.

\section{Results}

The experiments were done on the machine containing GTX Titan Xp GPU with 12 GB GDDR5X memory with the 384-bit bus at 5.705 GHz (547.68 GB/s), i7-4930K CPU, and 48 GiB DDR3 $4 \times 64$-bit memory at 1067 GHz (68.256 GB/s).
We used Linux operating system with CUDA compilation tools release 10.0 and Taichi language version 0.7.26.
Code profiling was done in NVIDIA Visual Profiler.

\subsection{Validation}

We have validated the code for the three standard cases: lid-driven cavity, flow through a channel and flow past a cylinder.
Due to simple boundary conditions and collision models, not all cases gave good agreement with physical models, but we used them as a method to validate code correctness.
All computations were done in single precision using the dense Taichi data layout.

\subsubsection{Lid-driven cavity}

The lid-driven cavity flow is a standard CFD benchmark, where the flow inside a square chamber is driven by a constant velocity at the chamber top lid. 
Depending on the Reynolds number, different vortex structures can be observed.
The characteristic length $L$ equals to the length of the chamber side ($L = n_y - 1$ in lattice units), the x-velocity of the top lid is the characteristic velocity $U$ (we used $U = 0.1$ in lattice units), and y-velocity of the top lid is 0.
The velocities at all other nodes are assumed to be zero. 
On the top wall, the constant velocity boundary condition was imposed. 
For other walls, we used the bounce-back boundary conditions.

\begin{figure}
	\centering
    \begin{subfigure}{0.35\textwidth}
    \centering
    \begin{tikzpicture}
        \node[inner sep=0pt] (cav3D) at (0,0) 
            {\includegraphics{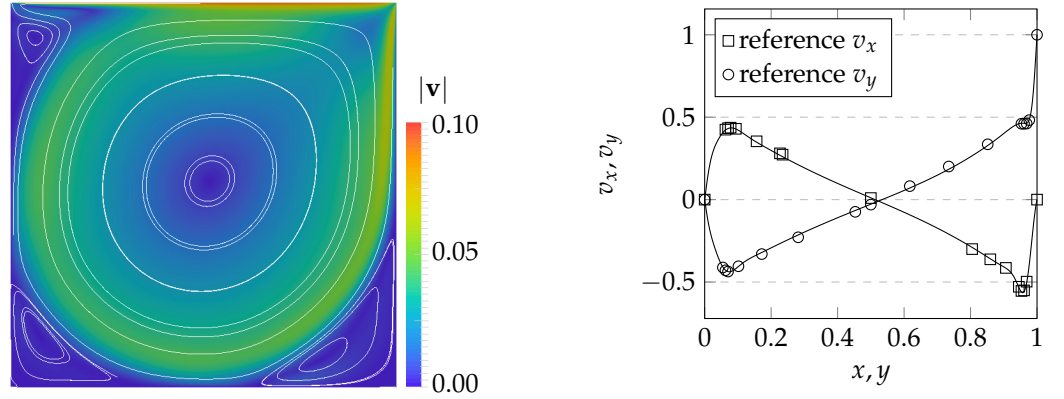}};
        \draw (2.5,1.4) node[anchor=west] {$\left| \vb{v} \right|$};
        \draw (2.7,0.95) node[anchor=west] {$0.10$};
        \draw (2.7,-0.75) node[anchor=west] {$0.05$};
        \draw (2.7,-2.5) node[anchor=west] {$0.00$};
    \end{tikzpicture}
    \end{subfigure}
    \hfill
    \begin{subfigure}{0.35\textwidth}
        \centering
        \begin{tikzpicture}
            \begin{axis}[
            	height=5.7cm,
                width=6cm,
                xmin=0, xmax=1,
                legend pos=north west,
                ymajorgrids=true,
                grid style=dashed,
                xlabel={$x, y$},
            	ylabel={$v_x, v_y$}
            	]
            
            \addplot[only marks, mark=square] 
            	table[x=x,y=v5000] {img/ghia1982.dat};
            \addlegendentry{reference $v_x$}

            \addplot[only marks, mark=o] 
            	table[x=y,y=u5000] {img/ghia1982.dat};
            \addlegendentry{reference $v_y$}

            \addplot[] 
            	table[x=coord, y=vx] {data/cav2D_Re_5000_3M_steps_2048x2048.dat};

            \addplot[] 
            	table[x=coord, y=vy] {data/cav2D_Re_5000_3M_steps_2048x2048.dat};
            	
            \end{axis}
        \end{tikzpicture}    
    \end{subfigure}
\caption{
	Lid-driven cavity results for $\mathrm{Re} = 5000$ and D2Q9 lattice with $2048 \times 2048$ resolution. 
	The left picture shows the magnitude of velocity and streamlines (denoted with white lines) after $3 \cdot 10^6$ time steps. 
	The right plot contains a comparison of velocity profiles with the reference data from \cite{Hig:Ghia1982}.
	The complete simulation took about 44 minutes on GPU.
}
    \label{fig:cav}
\end{figure}


\begin{specialtable}[H] 
\caption{
	Validated cavity simulations for different Reynolds numbers $\mathrm{Re}$.
	Time of computations is given for GTX Titan Xp GPU and includes the compilation of kernel (about 20 seconds) and a few dozens of saves of simulation state to disk.
\label{tab:cav2D_sim_params}}
\begin{tabular}{rcrc}
\toprule
\textbf{$\mathrm{Re}$}	& \textbf{Mesh resolution}  & \textbf{Number of time steps}  & \textbf{Computation time} \\
\midrule
  100                      &  $128 \times 128$         &      $12~000$                 & 29 seconds \\
 1000                      &  $256 \times 256$         &     $100~000$                 & 56 seconds \\
 3200                      & $1024 \times 1024$        &   $1~000~000$                 & ~4 minutes \\
 5000                      & $2048 \times 2048$        &   $3~000~000$                 & 44 minutes \\
10000                      & $4096 \times 4096$        &  $10~000~000$                 & 12.5 hours \\
\bottomrule
\end{tabular}
\end{specialtable}

The results were compared with data from \cite{Hig:Ghia1982} 
for different mesh resolutions $n_x \times n_y$, an example is shown in Figure \ref{fig:cav}.
To get the correct $\mathrm{Re}$ values, the fluid viscosity was computed according to Eqn. (\ref{eqn_Re}) as $\nu = U \cdot (n_y - 1) / \mathrm{Re}$.
A uniform fluid density $\rho = 1$ was imposed initially.
For some combinations of mesh resolution and Reynolds numbers (for small resolutions and high Re numbers, as well as for large resolutions and small Re numbers), we observed numerical instabilities.
Verified simulation settings are shown in Table \ref{tab:cav2D_sim_params}.
The simulation for $\mathrm{Re} = 10000$ was stopped before fully converged (mean squared error $\sum (v - v_{ref})^2 / n$  was at the order of $10^{-3}$, maximum relative error was about 30\% for $v_x$ at $y = 0.2813$) and stabilized (we could still observe small, fading swirling waves), but it was slowly approaching the reference velocity profiles.
We also observed that, for $\mathrm{Re} = 3200$, a single reference data point at $(y = 0.4531, v_x = 0.86636)$ was significantly different than others.

\subsubsection{Channel flow}

The flow through a channel can be solved analytically and is often used to validate the correctness of CFD solvers.
In the channel flow case, the fluid flow is analysed for a long channel with a radius $R$ (for 2D case the radius equals half of the channel height), and the initial conditions force the flow along the channel.
The stabilized flow should form a parabolic velocity profile 
\begin{equation}
	v_x = v_{max} \cdot \left( 1 - \left(\frac{r}{R} \right)^2 \right)
	\label{eqn:chan_prof}
\end{equation}
with maximum velocity $v_{max}$ at the centre of the channel ($r$ denotes the distance from the centre).
 
\begin{figure}[t]
	\centering
    \begin{subfigure}{0.35\textwidth}
    \centering
    \begin{tikzpicture}
        \node[inner sep=0pt] (cav3D) at (0,0) 
            {\includegraphics{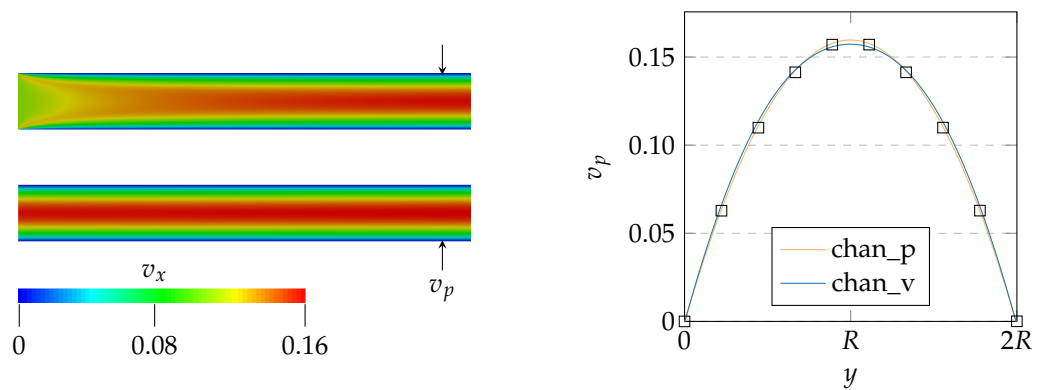}};
        \draw (-1.2,-0.9) node[anchor=north] {$v_x$};
				\draw (-3,-1.55) -- (-3,-1.85) node[anchor=north] {$0$};
				\draw (-1.2,-1.55) -- (-1.2,-1.85) node[anchor=north] {$0.08$};
				\draw (0.8,-1.55) -- (0.8,-1.85) node[anchor=north] {$0.16$};
				\draw [stealth-](2.625,-0.7) -- (2.625,-1.1) node[anchor=north] {$v_p$};
				\draw [stealth-](2.625,1.5) -- (2.625,1.9) node[anchor=north] {};
    \end{tikzpicture}
    \end{subfigure}
    \hfill
    \begin{subfigure}{0.35\textwidth}
        \centering
        \begin{tikzpicture}
            \begin{axis}[
            	height=5.7cm,
                width=6cm,
                xmin=0, xmax=1,
								xtick = {0, 0.5, 1},
								xticklabels = {0, $R$, $2R$},
								ymin = 0,
                ytick={0, 0.05, 0.1, 0.15},
                yticklabels={0, 0.05, 0.10, 0.15},
								legend style={at={(0.5,0.05)}, anchor=south},
                ymajorgrids=true,
                grid style=dashed,
                xlabel={$y$},
            		ylabel={$v_p$}
            	]

						\addplot[color6]
							table[x expr=\coordindex / 512, y index=0] {data/vel_prof_chan_p_15_16.dat};
							\addlegendentry{chan\_p}
						\addplot[color22]
							table[x expr=\coordindex / 512, y index=0] {data/vel_prof_chan_v_15_16.dat};
							\addlegendentry{chan\_v}
						\addplot[domain=0:1, samples=10, only marks, mark = square] {0.159 * (1 - ((x-0.5)/0.5)^2 )};

            \end{axis}
        \end{tikzpicture}    
    \end{subfigure}
\caption{
	Velocity $v_x$ for channel flows with constant inlet velocity $v_x = 0.1$ (left top picture) and constant pressure $\rho = 1.016$ (left bottom) together with velocity profiles $v_p$ (right plot).
	The profiles were taken at points $x = 15 \cdot R$ (marked with arrows), where $R$ is half the channel height.
	Squares denote reference values computed from Eqn. (\ref{eqn:chan_prof}) for arbitrarily chosen $v_{max} = 0.159$.
}
    \label{fig:channels}
\end{figure}

We tested two versions of channel flow differing with inlet boundary conditions.
Simulation parameters were set to get similar $v_{max}$ for both cases.
The first case, denoted as chan\_v, used the constant velocity $v_x = 0.1$ boundary condition at inlet.
For fluid nodes, initial density was set to $\rho (t_0) = 1.0$.
In the second case, chan\_p, the inlet boundary condition was set to constant pressure with $\rho = 1.016$, and the initial density for fluid nodes was set to $\rho (t_0) = 1.008$.
Both versions used channels containing $4096 \times 512$ nodes with bounce-back boundaries on top and bottom walls.
Outlet condition was set to constant pressure with $\rho = 1.0$, fluid viscosity was $\nu = 0.25$, initial velocity for fluid nodes was set to $\vb{v} (t_0) = \left[ 0,0 \right]$.
We calculated $10^6$ time steps what on GPU took about 10 minutes per case, including saves of simulation state every $10^4$ time steps.
The achieved computational kernel performance was 5.38 GLUPS, 387 GB/s for the single-precision version.

As can be seen in Figure \ref{fig:channels}, the obtained velocity profiles were close to parabolic.
The maximum values of velocities were slightly different ($v_{max} = 0.158$ for chan\_v and 0.160 for chan\_p).
Since we were using the quasi-compressible fluid model, then we were not able to achieve a complete agreement with theoretical models.

\subsubsection{Flow past cylinder}

\begin{figure}[t]
	\centering
    \begin{subfigure}{\textwidth}
			\includegraphics{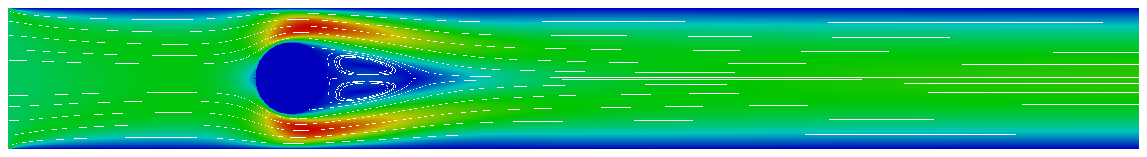}
			\caption{$\nu = 0.5$}
			\label{aa1}
		\end{subfigure}

		\vskip 4mm
    
		\begin{subfigure}{\textwidth}
			\includegraphics{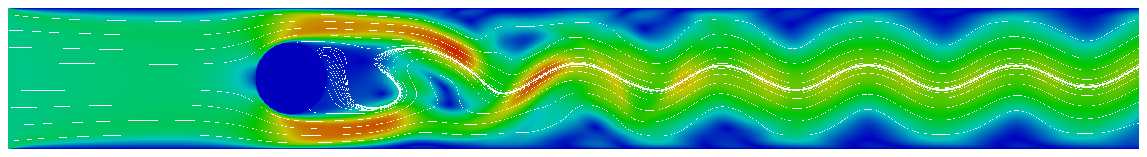}%
			\vskip 2mm
			\includegraphics{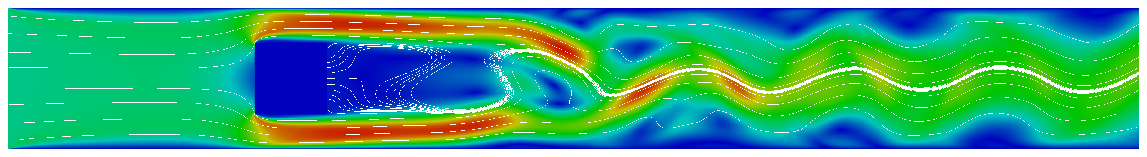}%
			\caption{$\nu = 0.1$}
			\label{aa3}
		\end{subfigure}

		\vskip 4mm
    
		\begin{flushleft}
    \begin{tikzpicture}
        \node[inner sep=0pt] (cav3D) at (0,0) 
            {\includegraphics{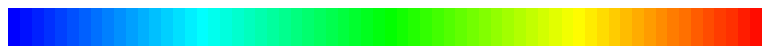}};
				\draw (-3.83,-0.2) -- (-3.83,-0.5) node[anchor=north] {$0$};
				\draw (-1.196,-0.2) -- (-1.196,-0.5) node[anchor=north] {$0.1$};
				\draw (1.439,-0.2) -- (1.439,-0.5) node[anchor=north] {$0.2$};
				\draw (3.81,-0.2) -- (3.81,-0.5) node[anchor=north] {$0.29$};
        \draw (-4.1,0) node[anchor=east] {$| \vb{v} |$};
    \end{tikzpicture}
		\end{flushleft}
\caption{
	Velocity magnitude $| \vb{v} |$ and streamlines (white) after $10^6$ time steps for flows past a cylinder at different viscosities $\nu$.
	The top picture shows a separation bubble comprising two symmetric and counter-rotating recirculation zones.
	Other images contain unsteady Kármán vortex street patterns.
	At the inlet, constant velocity is set to $v_x = 0.1$.
	Mesh resolution is $4096 \times 512$ nodes, sphere diameter and square edge are equal to half of the channel height.
}
    \label{fig:karman}
\end{figure}

Flows past stationary cylinders of different shapes are also one of the typical CFD problems.
For low Reynolds numbers, the flow is stationary.
With increasing $\mathrm{Re}$, the unsteady phenomenon called Kármán vortex street appears.

Our simulations were based on the chan\_v channel flow described above.
All parameters were identical except for viscosity which was changed to observe different flow patterns.
We used two often analysed, standard circular and square cylinders to find errors with handling boundary conditions at corner nodes.
The cylinders were placed at a distance from the inlet equal to two heights of the channel.
On the cylinder surface, we used the bounce-back boundary condition.
The simulation time and performance was similar to the chan\_v case.
The results are shown in Figure \ref{fig:karman} where typical behaviour can be observed.

\subsection{Performance}

\subsubsection{Memory bandwidth}

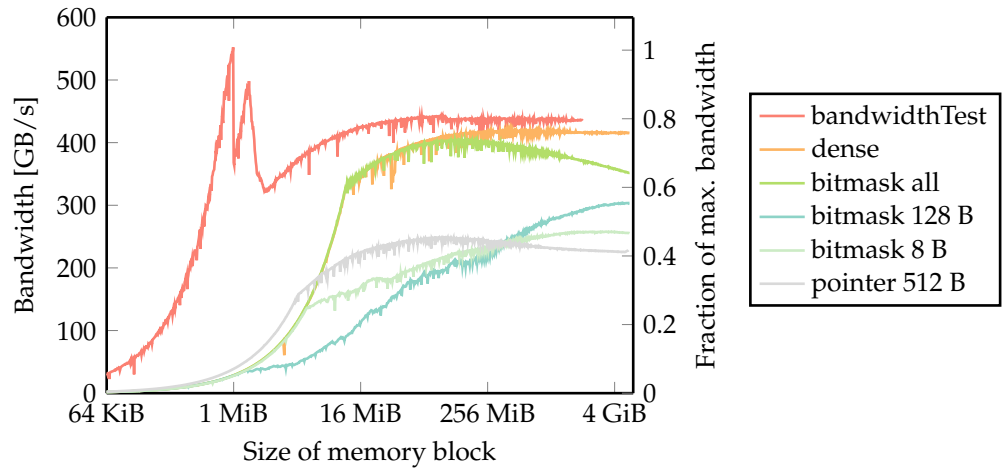
\begin{figure}
	\centering
        \begin{tikzpicture}
			\pgfplotsset{
				every axis legend/.append style={
					at={(1.22,0.8)},
					anchor=north west,
				},
			},
				\pgfplotsset{every axis/.append style={line width=1pt}
				},
            \begin{axis}[
								scale only axis,
            		height=5cm,
                width=7cm,
                ymin=0, ymax=600,
								axis y line*=left,
								ytick distance=100,
								xmin=65536, xmax=6442450944,
								xtick={65536, 1048576, 16777216, 268435456, 4294967296},
								xticklabels={64 KiB, 1 MiB, 16 MiB, 256 MiB, 4 GiB },
								xmode=log,
								log basis x=2,
								legend cell align=left,
								ymajorgrids=false,
                grid style=dashed,
                xlabel={Size of memory block},
            	  ylabel={Bandwidth [GB/s]}
            	]
            \addplot[color4] 
            	table[x index=0,y index=2] {data/bandwidthTest.dat};
            \addlegendentry{bandwidthTest}

            \addplot[color6] 
            	table[x index=0,y index=1] {data/ti_mem_bw_dense.log};
            \addlegendentry{dense}

            \addplot[color7] 
            	table[x index=0,y index=1] {data/ti_mem_bw_bitmask_all.log};
            \addlegendentry{bitmask all}

            \addplot[color1] 
            	table[x index=0,y index=1] {data/ti_mem_bw_bitmask_128.log};
            \addlegendentry{bitmask 128 B}

            \addplot[color11] 
            	table[x index=0,y index=1] {data/ti_mem_bw_bitmask.log};
            \addlegendentry{bitmask 8 B}

            \addplot[color9] 
            	table[x index=0,y index=1] {data/ti_mem_bw_pointer_512.log};
            \addlegendentry{pointer 512 B}

            \end{axis}
            \begin{axis}[
								scale only axis,
            		height=5cm,
                width=7cm,
								xmin=65536, xmax=6442450944,
								ymin=0, ymax=1.09553023663453111305,
								xmode=log,
								log basis x=2,
                ymajorgrids=false,
                grid style=dashed,
            	  ylabel={Fraction of max. bandwidth},
								axis x line=none,
								axis y line*=right,
            	]
							\end{axis}
            
        \end{tikzpicture}    
 \caption{
	 Bandwidth comparison of CUDA \emph{bandwidthTest} and Taichi kernels copying linear memory.
	 Numbers after "bitmask" and "pointer" denote the size of the data block assigned to a single bitmask/pointer.
	 }
 \label{fig:mem_bw}
\end{figure}

Before analysis of the implemented kernel, we first measured available memory bandwidth during a simple copy of data between 1-dimensional arrays for different data layouts implemented in Taichi.
The results are shown in Figure \ref{fig:mem_bw}.
Notice that we use both SI (k $=10^3$, M $=10^6$, G $=10^9$) and binary (Ki $=2^{10}$, Mi $= 2^{20}$, Gi $= 2^{30}$) prefixes.
As an approximate reference, we run the NVidia \emph{bandwidthTest} utility.
Internally, \emph{bandwidthTest} measures calls to \emph{cudaMemcpy} method, but it should be noted that this utility does not use preliminary "warm up" of the measured code.
Also, the \emph{bandwidthTest} has no support for size arguments larger than $2^{31}$ bytes, and the results are displayed in MiB/s and must be scaled.
We have not investigated the strange behaviour observed around size 1 MiB, but we believe that the values starting from a range of single MiB are reasonable. 

For Taichi, we prepared a simple kernel copying data from a 1-dimensional array to the other.
The number of threads per thread block was explicitly set to 512 since it resulted in high average performance.
Time duration measurements were done using Python \emph{time.perf\_counter} for 100 kernel calls (as in \emph{bandwidthTest}). 
Additionally, before each measurement, we initially called the measured kernel five times to force runtime compilation and warm up the whole system.

As can be seen in Figure \ref{fig:mem_bw}, the memory bandwidth measured for Taichi kernels strongly depends on the used data layout. 
For the large size of transferred data, the maximum transfer (419.7 GB/s for 500 MB block size) was observed for the standard dense layout. 
In the gigabyte range, the average bandwidth was 416 GB/s which is only 5\% lower than 436 GB/s reported by \emph{bandwidthTest}.

When the bitmasked layout is used, the memory bandwidth strongly depends on the amount of data masked by a single bit.
For a single masking bit per 8 bytes of data, the maximum bandwidth was 258 GB/s (at about 4 GiB block size).
When a single bit was used to mask the whole block of data and small data blocks were transferred, the measured bandwidth was similar to that for dense layout.
However, starting from about 200 MiB block size, the bandwidth slowly dropped to 352 GB/s for a 5.41 GiB block size.
For many different sizes of bytes per single bitmask, e.g. 64 KiB, the measured bandwidth dropped even to less than 30 GB/s.
For fine-grained bitmasking, 
the highest bandwidth (up to 304 GB/s when few GiB of data were transferred) was observed when 128-byte blocks were masked.

The pointer layout has the lowest maximum measured bandwidth, although it is less dependent on the size of the block of data assigned to a single pointer than the bitmask layout.
For a single pointer per 8 bytes of data, the bandwidth was below 20 GB/s and, additionally, we had to significantly decrease the amount of allocated memory.
With the increasing size of the pointed data block, the bandwidth steadily increased, achieving 247 GB/s at 124 MiB transferred data block for a single pointer per 512 bytes of data, as shown in Figure \ref{fig:mem_bw}.
Then, with increasing the amount of data per single pointer, the maximum bandwidths stayed in the 230-250 GB/s range for up to 16 MiB of data per pointer.
After this limit, the bandwidth dropped to 144 GB/s when a single pointer was used for a block containing 1 GiB of data.
We can also observe that the pointer layout has different overheads than other data layouts - for the small size of the data block, the performance plot for the pointer layout has a different shape than for other layouts.

The data from Figure \ref{fig:mem_bw} allow us to draw a general conclusion that for dense data layouts, the Taichi language brings in a small, usually negligible overhead.
However, sparse layouts (bitmask and pointer) reduce available bandwidth by at least about 40\% for fine-grained resolution.
It should be noticed, that for all data layouts available in Taichi we were trying to find parameters giving the highest bandwidth, although we did not do the exhausting measurements for all available combinations.
We could then miss some significantly better settings, although it seems unlikely.

\subsubsection{LBM performance}

The performance $P$ of LBM implementations is often measured in \emph{lattice updates per second} (LUPS), which define the number of processed nodes per unit of time.
However, direct comparison of $P_{LUPS}$ values for different implementations is difficult since the amount of processed data per node depends on both the lattice arrangement and the data type, usually either single (32 bits) or double (64 bits) precision floating-point number.
Since LBM implementations are usually bandwidth-bound on available machines, in this work, we then define LBM performance as a theoretical, minimal bandwidth $P_B$ required to achieve given LUPS for specific lattice arrangements and data types.

Let $s_d$ denote the size (number of bytes) required to store a single number ($s_d \in \{4, 8 \}$ for single- and double-precision floating-point numbers).
Assuming that, in an ideal case, processing of a lattice node requires only read and write of all $q$ functions $f_i$, the minimum amount of transferred data per single node is $B_{node} = 2 \cdot q \cdot s_d$ bytes.
The LBM implementation performance can be then defined as $P_B = B_{node} \cdot P_{LUPS}$.
We can also define a theoretical bandwidth utilisation $U_B = P_B / B_{peak}$, where $B_{peak}$ denotes the maximum theoretical memory bandwidth of a given machine.
For the GTX Titan Xp GPU, $U_B = C \cdot P_{GLUPS}$, where $C \in \{ 0.131, 0.263 \}$ for single and double precision, respectively, and $P_{GLUPS}$ is performance in $10^9$ LUPS.
The values of $U_B$ can be then compared for different machines showing how much room for potential improvements is still available.
However, the $U_B$ coefficient does not take into account the additional limitations of the machine that prevent it from achieving full memory bandwidth. 

The performance of the LBM kernel was measured for dense and sparse geometries, different data structures, and single and double precision numbers.
Results for sparse data structures for dense geometry can be treated as an estimation of introduced overheads compared to the dense memory layout.
Measurements were done using Python \emph{time.perf\_counter} for 1000 kernel calls.
Before measurements, each kernel was called 100 times to force the runtime compilation and warm up the whole system.
For sparse data layouts, we turned on the experimental \emph{async\_mode} available in Taichi that disables analysis of geometry sparsity before each kernel call because, in our case, the geometry is static during computations.
This mode allowed to increase performance, but we observed sporadical problems with stability.

\paragraph{Performance for dense geometries}

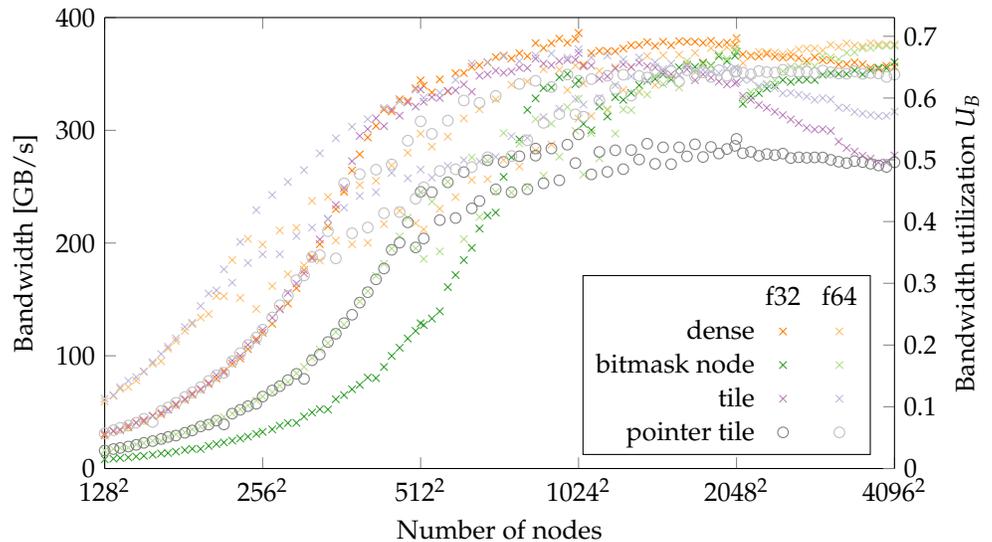
\begin{figure}
	\centering
        \begin{tikzpicture}
						\pgfplotsset{
								legend image with text/.style={
										legend image code/.code={%
												\node[anchor=center] at (0.3cm,0cm) {#1};
										}
								},
						}
            \begin{axis}[
								scale only axis,
            		height=6cm,
                width=10.5cm,
                xmin=128, xmax=4096,
                ymin=0, ymax=400,
								axis y line*=left,
								ytick distance=100,
								xtick={128, 256, 512, 1024, 2048, 4096},
								xticklabels={$128^2$, $256^2$, $512^2$, $1024^2$, $2048^2$, $4096^2$},
								xmode=log,
								log basis x=2,
            		legend columns=2,
            		legend style={
            		    legend cell align=right,
            		},
            		legend pos=south east,
								legend plot pos=right,
								ymajorgrids=false,
                xlabel={Number of nodes},
            	  ylabel={Bandwidth [GB/s]}
            	]

							\addlegendimage{legend image with text=$\mathrm{f32}$}
							\addlegendentry{}
							\addlegendimage{legend image with text=$\mathrm{f64}$}
							\addlegendentry{}

            \addplot[color28, only marks, mark=x] 
            	table[x index=0,y index=2] {data/perf_cav2D_Titan_Xp_128_4096_dense_f32.dat};
            \addlegendentry{dense}
            \addplot[color27, only marks, mark=x] 
            	table[x index=0,y index=2] {data/perf_cav2D_Titan_Xp_128_4096_dense_f64.dat};
            \addlegendentry{}

            \addplot[color24, only marks, mark=x] 
            	table[x index=0,y index=2] {data/perf_cav2D_Titan_Xp_128_4096_bitmask_node_f32.dat};
            \addlegendentry{bitmask node}
            \addplot[color23, only marks, mark=x] 
            	table[x index=0,y index=2] {data/perf_cav2D_Titan_Xp_128_4096_bitmask_node_f64.dat};
            \addlegendentry{}

            \addplot[color10, only marks, mark=x] 
            	table[x index=0,y index=2] {data/perf_cav2D_Titan_Xp_128_4096_tile_f32.dat};
            \addlegendentry{tile}
            \addplot[color3, only marks, mark=x] 
            	table[x index=0,y index=2] {data/perf_cav2D_Titan_Xp_128_4096_tile_f64.dat};
            \addlegendentry{}

            \addplot[gray50, only marks, mark=o] 
            	table[x index=0,y index=2] {data/perf_cav2D_Titan_Xp_128_4096_pointer_tile_f32.dat};
            \addlegendentry{pointer tile}
            \addplot[gray75, only marks, mark=o] 
            	table[x index=0,y index=2] {data/perf_cav2D_Titan_Xp_128_4096_pointer_tile_f64.dat};
            \addlegendentry{}

            \end{axis}
            \begin{axis}[
								scale only axis,
            		height=6cm,
                width=10.5cm,
                xmin=128, xmax=4096,
								xmode=log,
								axis x line=none,
								log basis x=2,
								axis y line*=right,
								ymin=0, ymax=0.73034015592762329054,
								ytick distance=0.1,
                ymajorgrids=false,
            	  ylabel={Bandwidth utilization $U_B$},
            	]
							\end{axis}
            
        \end{tikzpicture}    
 \caption{
		Performance of cavity GPU simulation for different mesh sizes, memory layouts and data types (f32 and f64 denote single- and double-precision floating-point numbers, respectively).
	 }
 \label{fig:cav_dense_perf}
\end{figure}

As an example of dense geometry, we used the cavity case at different mesh resolutions.
Measurements for each data point require more than 20 seconds because of kernel compilation.
Thus, we limited the number of checked mesh resolutions to about 100 on a logarithmic scale.
To keep results comparable, all measurements for different data types and layouts are for the same set of geometry resolutions.
The obtained performance is shown in Figure \ref{fig:cav_dense_perf} and Table \ref{tab:perfCav}.

We compared performance for four different data layouts.
The dense layout is a standard, multidimensional array containing all PDFs for all nodes.
As excepted, this version offers the highest performance to 70\% of peak GPU memory bandwidth.
This result is close to the best-reported values for highly optimised codes from \cite{Mawson:Mem2014, Januszewski:Sail2014} and is consistent with performance reported for the same hardware in \cite{Tomczak:Ane2019}.
During code profiling, we observed that loads of all PDFs (excluding $f_0, f_2$ and $f_5$) cause uncoalesced transactions because, as shown in \cite{Januszewski:Sail2014}, neighbour $f_i$ values are shifted in memory by one position.
We found no method to correct this behaviour - the typical technique is the usage of shared memory but the Taichi language offers no such feature.

The bitmasked data structures can be used in many ways.
At first, we applied a single bitmask per each $f_i$ function because, due to the structure-of-arrays data layout, we were not able to apply a single bitmask per a whole lattice node.
The observed performance dropped more than twice compared to the dense layout - maximum bandwidth was at the level of 140 GB/s for the single-precision version.


\begin{specialtable}[t] 
\caption{
	Performance of cavity simulations for different data layouts.
	Performance $P_{LUPS}$ is given in GLUPS, $P_B$ in GB/s.
	Column \textbf{Mesh} contains mesh size for which maximum performance was observed.
	Average performance is computed for meshes containing at least $1024^2$ nodes.
\label{tab:perfCav}}
\begin{tabular}{lccccccccc}
\toprule
\multirow{2}{*}{\textbf{Layout}} & & \multicolumn{4}{c}{\textbf{Maximum}} & & \multicolumn{3}{c}{\textbf{Average}} \\
										  \cline{3-6}                                       \cline{8-10} 
                     & & \textbf{Mesh} & $P_{LUPS}$  &  $P_B$  &  $U_B$   & & $P_{LUPS}$  &  $P_B$  &  $U_B$  \\
\midrule                                                                 
f32 dense             & &   $1024^2$    &    5.37		 &   386   &  0.705   & &    5.12		 &   369   &  0.674  \\
f32 bitmask node      & &   $2048^2$    &  	 5.13    &   370   &  0.675   & &    4.78    &   344   &  0.629  \\
f32 tile              & &   $1024^2$    &    5.12    &   369   &  0.673   & &    4.53    &   326   &  0.596  \\
f32 pointer tile      & &   $1024^2$    &    4.12    &   296   &  0.541   & &    3.85    &   277   &  0.506  \\
                      & &               &            &         &          & &            &         &         \\
f64 dense             & &   $2048^2$    &    2.65 	 &   381   &  0.696   & &    2.51		 &   362   &  0.661  \\
f64 bitmask node      & &   $3955^2$    &    2.60    &   375   &  0.685   & &    2.40    &   346   &  0.631  \\
f64 tile              & &   $1024^2$    &    2.58    &   372   &  0.679   & &    2.37    &   341   &  0.622  \\
f64 pointer tile      & &   $2509^2$    &    2.46    &   354   &  0.646   & &    2.42    &   348   &  0.635  \\
\bottomrule
\end{tabular}
\end{specialtable}

However, the bitmask layout does not save memory and serves only as a convenient way to skip computations for non-existent data.
The reasonable way is then to use bitmasks only for a field containing encoded node type.
This layout enables for simple management of sparse geometries and is marked as "bitmask node" on performance plots.
As can be seen in Figure \ref{fig:cav_dense_perf} and Table \ref{tab:perfCav}, when a single bitmask is used per whole data of a single lattice node, the performance loss is less than 10\% compared to the dense layout for geometries containing at least $10^6$ nodes.
For smaller geometries, the bitmask layout has low performance.
An additional advantage of this approach is that, due to the fine-grained masking of single nodes, only valid nodes are processed, even for very complex geometries.

The pointer layout available in Taichi can also be used in many ways, but the applied method should allow storing in memory only values used during computations.
We then used a single pointer per \emph{tile} containing data for $16^2$ neighbour nodes.
The resulting data layout is denoted as "pointer tile".
Additionally, we measured performance for the "tile" layout defined as a dense array of tiles without an additional layer of pointers.

The measured performance of the tile layout was similar to the dense layout, but only for geometries with less than $2048^2$ nodes.
After this limit, the achieved bandwidth utilisation dropped even below 0.5 for single-precision data and the largest geometry.
We observed two issues appearing for the tile data layout.
First, the dimension of the CUDA thread block had to be reduced to the number of nodes per tile - in our case from 512 to 256 threads per block.
For the dense layout, such a change of thread block size decreased bandwidth from 360 to 338 GB/s for the cavity $4096^2$ case and single-precision data.
Next, the Taichi has no support for low-level optimisations presented in \cite{Tomczak:Ane2019, Tomczak:Spa2018}, e.g. usage of shared memory, warp level programming, and LBM-optimized index calculations inside a tile.
For example, we tried to store $f_1$ and $f_3$ functions using column-major order, but we encountered different behaviour than described in Taichi documentation.
It is also probable that the observed decrease in performance may be caused by other, undetected yet reasons.

For the pointer tile layout, the performance is slightly surprising.
When double-precision data is used, then the performance stays high and steady even for the largest geometries, despite the drop-out observed for tile layout.
On the other hand, performance for single-precision is also almost constant but at the low level ($U_B = 0.5$) given by the minimum value observed for tile layout and the largest geometry.
We have not found the cause of such behaviour yet and only found that the code handling pointers in the kernel is quite complex and significantly increases register pressure - the kernel required 70 registers what decreased the theoretical occupancy to 37.5\%.

The presented data shows that we achieved high performance for each of the presented layouts despite the low bandwidth observed during simple data copy of sparse layouts.
Only the tile-based layouts have lower performance, and for large geometries only (except the pointer tile layout for single-precision data).
However, the measured performance was erratic and strongly dependent on geometry size for some of the analysed layouts, for example, the bitmask node layout, single-precision data,  and geometries containing between $1024^2$ and $2048^2$ nodes.

\paragraph{Performance for sparse geometries}

Performance for sparse geometries is measured in the same way as for dense ones, but only non-solid nodes are taken into account in the performance calculation.
Thus, for each sparse geometry, we define its \emph{porosity}
\begin{equation}
	\phi = \frac{n_{\mathrm{non-solid~nodes}}}{n_{\mathrm{all~nodes}}}
\end{equation}
which determines what proportion of all nodes is involved in the LBM calculations.
We treat all non-solid nodes as computational because LBM implementation is bandwidth-bound on our system and, even for the bounce-back boundary nodes, which do not require computations, we need to read and write $f_i$ functions from/to memory. 

The performance for sparse geometries was measured for square geometries with $4096^2$ nodes and different porosities $\phi \geq 0.1$.
The large geometry size was chosen to keep at least $10^6$ non-solid nodes, even for the lowest porosity.
To obtain the required $phi$, the geometry was filled with solid, circle obstacles.
We used two different arrangements of obstacles: a regular array and a random placement.
The regular array contains a mesh of $8 \times 8$ circles, which radius depends on the required $\phi$.
We did not use regular arrays for $\phi < 0.3$ because, in such cases, all geometry walls are filled with solid nodes from overlapping circles.
For the random placement, the geometry was filled with randomly placed circles.
The radii of circles were also randomly chosen from $r \in \left[8, 256\right]$ nodes.

To estimate overheads for computations on sparse geometries, we define the \emph{sparse computational efficiency}
\begin{equation}
	\eta_P = \frac{P (\mathrm{sparse~geometry})}{P (\mathrm{dense~geometry})}
	\label{eqn_eff}
	,
\end{equation}
where $P (\mathrm{sparse~geometry})$ and $P (\mathrm{dense~geometry})$ denote measured performances for the same data layout and geometry size.
It should be noted that $\eta_P$ does not take into account that the number of computational nodes in the sparse geometry is lower than in the dense one what may have an additional impact on performance.
However, for dense data layouts, the memory is allocated for all nodes regardless of a node type.
Thus, the definition in Eqn. (\ref{eqn_eff}) seems reasonable.

\begin{figure}[t]
	\centering
    \begin{subfigure}{0.35\textwidth}
    \centering
		\includegraphics{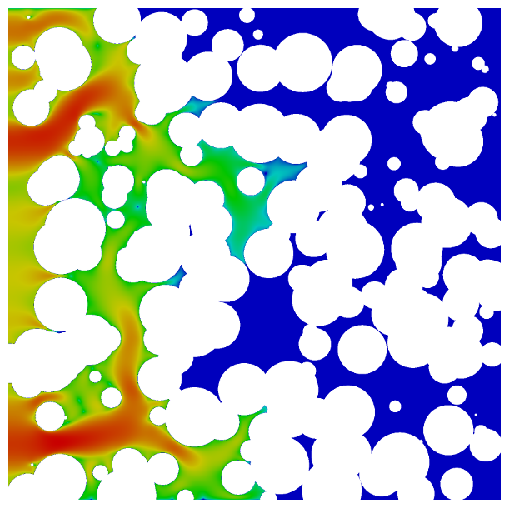}
    \end{subfigure}
    \hfill
    \begin{subfigure}{0.35\textwidth}
        \centering
		\includegraphics{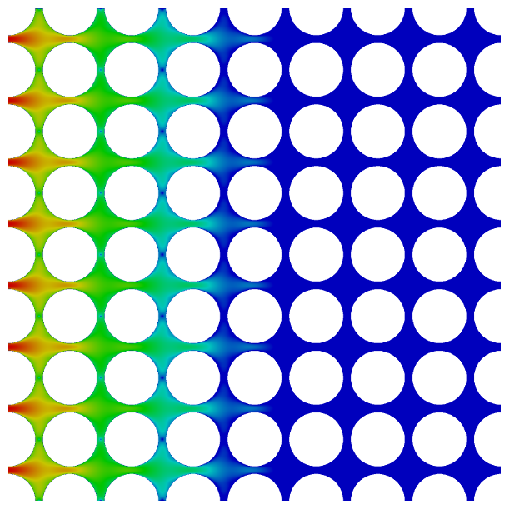}
    \end{subfigure}
\caption{
	Examples of sparse geometries with $4096^2$ nodes and porosity $\phi = 0.4$.
	Pressure boundary conditions were set on inlet (left wall, $\rho = 1.016$) and outlet (right wall, $\rho = 1.0$), bounce back on top and bottom walls and on borders of obstacles, viscosity $\nu = 0.5$.
	Colours correspond to velocity magnitude after $10^6$ time steps, a logarithmic scale is used.
	Calculations took about half an hour per geometry.
}
    \label{fig:sparse}
\end{figure}

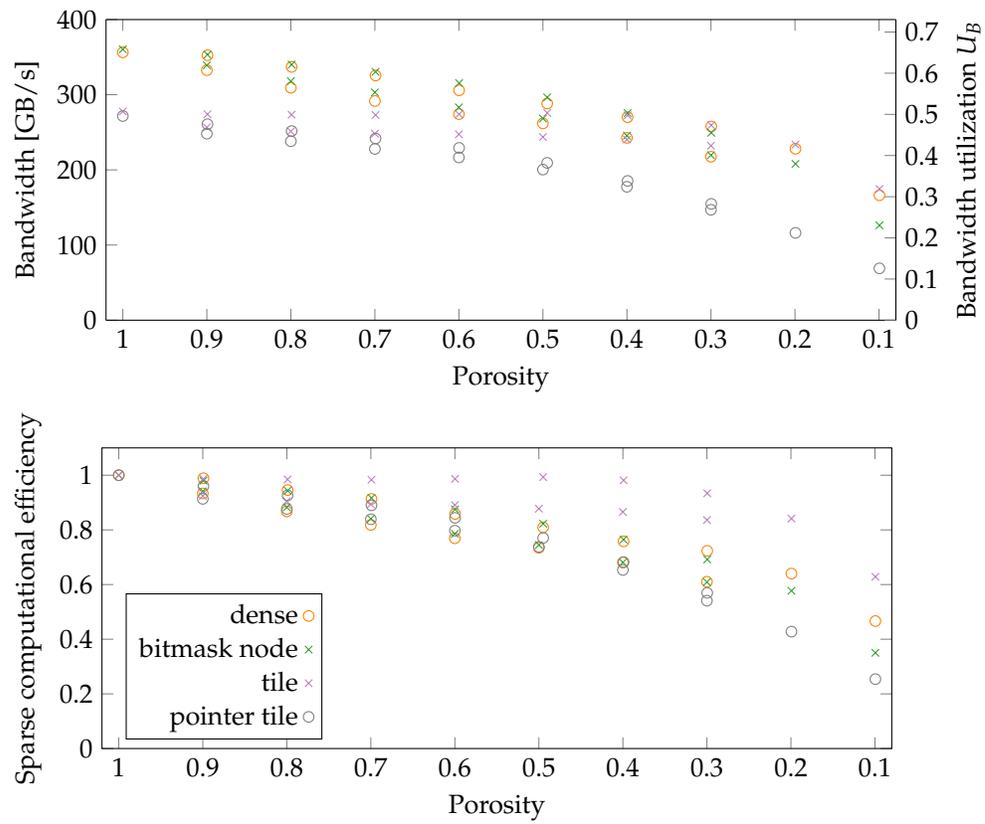
\begin{figure}[t]
	\centering
		\begin{subfigure}{\textwidth}
        \begin{tikzpicture}
            \begin{axis}[
								scale only axis,
            		height=4cm,
                width=10.5cm,
                xmin=0.08, xmax=1.02,
								x dir= reverse,
                ymin=0, ymax=400,
								axis y line*=left,
								ytick distance=100,
            		legend style={
            		    legend cell align=right,
            		},
            		legend pos=south west,
								legend plot pos=right,
								ymajorgrids=false,
                xlabel={Porosity},
            	  ylabel={Bandwidth [GB/s]}
            	]

            \addplot[color28, only marks, mark=o, /pgf/number format/read comma as period] 
            	table[x index=0,y=dense] {data/sparse_performance.csv};

            \addplot[color24, only marks, mark=x, /pgf/number format/read comma as period] 
            	table[x index=0,y=bitmasknode] {data/sparse_performance.csv};

            \addplot[color10, only marks, mark=x, /pgf/number format/read comma as period] 
            	table[x=porosity, y=tile] {data/sparse_performance.csv};

            \addplot[gray50, only marks, mark=o, /pgf/number format/read comma as period] 
            	table[x index=0,y=pointertile] {data/sparse_performance.csv};

            \end{axis}
            \begin{axis}[
								scale only axis,
            		height=4cm,
                width=10.5cm,
                xmin=128, xmax=4096,
								xmode=log,
								axis x line=none,
								log basis x=2,
								axis y line*=right,
								ymin=0, ymax=0.73034015592762329054,
								ytick distance=0.1,
                ymajorgrids=false,
            	  ylabel={Bandwidth utilization $U_B$},
            	]
							\end{axis}
            
        \end{tikzpicture}    
				\end{subfigure}

				\vfill

			\begin{subfigure}{\textwidth}
        \begin{tikzpicture}
            \begin{axis}[
								scale only axis,
            		height=4cm,
                width=10.5cm,
                xmin=0.08, xmax=1.02,
								x dir= reverse,
                ymin=0, ymax=1.1,
								ytick distance=0.2,
            		legend style={
            		    legend cell align=right,
            		},
            		legend pos=south west,
								legend plot pos=right,
								ymajorgrids=false,
                xlabel={Porosity},
            	  ylabel={Sparse computational efficiency}
            	]

            \addplot[color28, only marks, mark=o, /pgf/number format/read comma as period] 
            	table[x index=0,y=effdense] {data/sparse_performance.csv};
            \addlegendentry{dense}

            \addplot[color24, only marks, mark=x, /pgf/number format/read comma as period] 
            	table[x index=0,y=effbitmasknode] {data/sparse_performance.csv};
            \addlegendentry{bitmask node}

            \addplot[color10, only marks, mark=x, /pgf/number format/read comma as period] 
            	table[x=porosity, y=efftile] {data/sparse_performance.csv};
            \addlegendentry{tile}

            \addplot[gray50, only marks, mark=o, /pgf/number format/read comma as period] 
            	table[x index=0,y=effpointertile] {data/sparse_performance.csv};
            \addlegendentry{pointer tile}

            \end{axis}
        \end{tikzpicture}   
			\end{subfigure} 
 \caption{
		Performance of GPU simulations (top) and sparse performance efficiency (bottom) for single-precision numbers, different data layouts, and sparse geometries with $4096^2$ nodes and different porosities.
		For a given data layout, the lower sequence of points shows performance for regularly placed circles, and the upper points correspond to performance for randomly placed circles.
	 }
 \label{fig:sparse_perf}
\end{figure}

The obtained performance results are shown in Figure \ref{fig:sparse_perf}.
As can be seen, for all layouts but tile, the performance gradually drops with porosity.
This performance decrease shows that the geometry sparsity introduces overheads, for example, redundant memory traffic caused by neighbouring solid and non-solid nodes, which are visible even after using data layouts (e.g. bitmasked) eliminating explicit operations for solid nodes.
It should also be noted that the performance depends not only on porosity but also on the placement of solid and non-solid nodes.
For randomly placed obstacles, the measured performance was slightly higher than for the regular placement, because randomly placed circles formed large, solid areas that minimised the interlacing in memory of the data for solid and non-solid nodes.
Only for the tile layout, the performance was practically constant for porosities $\phi \geq 0.4$, which suggests that for the tile layout, other factors limit performance, as shown in Figure \ref{fig:cav_dense_perf} for large geometry.

The lowest performance was observed for the pointer tile layout.
For porosity 0.1, the performance dropped to 25\% of performance for dense geometry. 
However, in contrast to other layouts, performance for the pointer tile layout has the lowest differences between geometries with regularly and randomly placed obstacles.
It may suggest additional overheads not connected with a sparsity of geometry, although more detailed research is needed.

\subsubsection{Memory usage}

Memory usage is difficult to measure for the Taichi language since it internally preallocates and manages GPU memory.
However, we observed that, for the dense layout, the maximum size of allocated arrays was very similar for raw CUDA and Taichi implementations (two arrays containing about $720 \cdot 2^{20}$ 64-bit elements).
After exceeding this limit, we observed excessive page faults and a significant performance decrease.
When we turned off the unified memory support, then memory allocation errors appeared sporadically.
Although we did not conduct an in-depth analysis, we did not find any serious problems, thus it seems that Taichi effectively manages memory and introduces minimal overhead only.

\section{Conclusions}

In this work, we presented the implementation of the lattice-Boltzmann solver in Taichi, the interpreted, data-oriented language which decouples computations and data arrangement in memory.
We showed that, although sparse data layouts provided by Taichi bring significant overheads when used on a fine-grained level during simple data copying, it was possible to a design high-performance code of the lattice-Boltzmann solver for non-trivial cases.
Four data layouts have been tested: the dense layout that is a simple, multidimensional array; the tile layout with data arranged in square tiles containing $16^2$ neighbouring nodes; the bitmask node layout where single nodes could be masked; and the pointer tile layout allowing to allocate memory per single tiles but at the cost of additional, indirect addressing.
An additional advantage of the presented solution is short and simple code (about 1000 source lines) which is freely available in the hope that it can be a practical basis for further experiments due to its low complexity and high performance.

The obtained performance is comparable to the best existing implementations but strongly dependent on used data layouts.
For the dense layout, we achieved up to about 70\% of peak memory bandwidth available on GTX Titan Xp GPU, which corresponds to 5.37 GLUPS for single- and 2.65 GLUPS for double-precision computations.
Slightly lower performance, but still up to over 67\% of peak memory bandwidth, was observed for other dense data layouts, the tile and the bitmask node.
The lowest performance, up to 54\% of the peak, was obtained for the pointer tile layout and single-precision computations.
Thus, Taichi implementations of LBM for dense geometries should have the highest performance with the simple, dense layout.

For sparse geometries, the layout resulting in the highest performance depends on geometry sparsity.
Best performance for low porosities $\phi \leq 0.3$ was observed for the tile and the dense layouts, although it was still about two times slower than for dense geometries.
Geometries with higher porosities were processed the fastest with the bitmask node and the dense layouts.
The pointer tile layout has the lowest performance (about two times lower than the tile layout for porosity 0.1) but, in contrary to the other layouts, allows to save memory by skipping data for some solid nodes.

Apart from data layouts available in Taichi, also other factors may have a significant impact on the code performance, which sometimes was difficult to predict.
In the presented measurements, we observed that both geometry size and placement of solid and non-solid nodes has a visible impact on performance.
Full performance requires geometries containing at least $10^6$ nodes, but such behaviour was also reported in other papers about LBM implementations on GPU.
However, we have observed uncommon performance drop for the tile layout and large geometries, and significant performance limitations for the pointer tile layout for single-precision data.
We have not found the reason yet and believe that significantly more thorough studies are needed to analyse overheads introduced by the Taichi language and its internal architecture.

The source code in Taichi is clean and concise, but we observed a few limitations.
For example, we have not found a way to pass a static argument to function and use it as a compile-time constant index, thus we had to inline some operations manually to enable compile-time optimisations.
Also, it is difficult to control register usage per kernel, and setting the number of CUDA threads per block is limited.
For async mode, we sometimes observed problems with stability and code profiling by NVIDIA Visual Profiler.

One of the issues was also the long compilation time.
Although an additional time required for runtime compilation is typical for interpreted languages, it should be reported, especially that the presented code takes more than 20 seconds to compile.
Comparing this with a few seconds needed to obtain the stationary solution for small meshes (up to $256^2$ nodes), the kernel compilation enlarges the simulation time by order of magnitude.
We suspect that maybe some form of precompiled kernels could significantly decrease the time for short simulations.
It should be noted that we used only one simple collision model and a reduced set of boundary conditions.
For more complex computational models or universal kernels with support for many different collision models, the time required to generate kernel code can be longer.

Future work includes searching for methods that allow using other optimisation techniques used for LBM implementations.
We are also planning the implementation of more complex collision models, boundary conditions, and support for three-dimensional geometries, although this may require the new design of code due to a larger amount of data per node which can increase register pressure.

\vspace{6pt} 

\funding{This research received no external funding.}

\institutionalreview{Not applicable.}

\informedconsent{Not applicable.}

\dataavailability{The source code presented in this study is openly available in Github at \url{https://github.com/tadeusz-tomczak/tilb}.
} 

\acknowledgments{
	The authors gratefully acknowledge the support from NVIDIA Corporation for providing them the Titan Xp GPU used in this research.
}

\conflictsofinterest{The authors declare no conflict of interest.}

\abbreviations{Abbreviations}{The following abbreviations are used in this manuscript:\\

\noindent 
\begin{tabular}{@{}ll}
BGK & Bhatnagar–Gross–Krook operator \\
CFD & Computational fluid dynamics \\
CPU & Central processing unit\\
CUDA & Compute Unified Device Architecture \\
DRAM & Dynamic random-access memory \\
GPU & Graphics processing unit\\
GUI & Graphical user interface \\
ILP & Instruction-level parallelism \\
LBM & Lattice-Boltzmann method\\
LUPS & Lattice updates per second \\
PDF & Particle distribution function \\
SI & Système international (d'unités)\\
SIMD & Single instruction, multiple data \\
SNode & Structural node \\
\end{tabular}}

\end{paracol}
\reftitle{References}

\end{document}

%% file: img/la_d2q9.tex
\begingroup%
  \makeatletter%
  \providecommand\color[2][]{%
    \errmessage{(Inkscape) Color is used for the text in Inkscape, but the package 'color.sty' is not loaded}%
    \renewcommand\color[2][]{}%
  }%
  \providecommand\transparent[1]{%
    \errmessage{(Inkscape) Transparency is used (non-zero) for the text in Inkscape, but the package 'transparent.sty' is not loaded}%
    \renewcommand\transparent[1]{}%
  }%
  \providecommand\rotatebox[2]{#2}%
  \ifx\svgwidth\undefined%
    \setlength{\unitlength}{141.73228346bp}%
    \ifx\svgscale\undefined%
      \relax%
    \else%
      \setlength{\unitlength}{\unitlength * \real{\svgscale}}%
    \fi%
  \else%
    \setlength{\unitlength}{\svgwidth}%
  \fi%
  \global\let\svgwidth\undefined%
  \global\let\svgscale\undefined%
  \makeatother%
  \begin{picture}(1,0.7)%
    \put(0,0){\includegraphics[width=\unitlength]{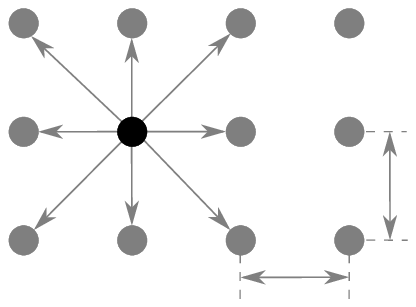}}%
    \put(0.48,0.46){\color[rgb]{0,0,0}\makebox(0,0)[lb]{\smash{1}}}%
    \put(0.1,0.46){\color[rgb]{0,0,0}\makebox(0,0)[lb]{\smash{3}}}%
    \put(0.34,0.6){\color[rgb]{0,0,0}\makebox(0,0)[lb]{\smash{2}}}%
    \put(0.34,0.2){\color[rgb]{0,0,0}\makebox(0,0)[lb]{\smash{4}}}%
    \put(0.55,0.6){\color[rgb]{0,0,0}\makebox(0,0)[lb]{\smash{5}}}%
    \put(0.14,0.6){\color[rgb]{0,0,0}\makebox(0,0)[lb]{\smash{6}}}%
    \put(0.1,0.25){\color[rgb]{0,0,0}\makebox(0,0)[lb]{\smash{7}}}%
    \put(0.5,0.25){\color[rgb]{0,0,0}\makebox(0,0)[lb]{\smash{8}}}%
    \put(0.97,0.27){\color[rgb]{0,0,0}\makebox(0,0)[lb]{\smash{$\delta x$}}}%
    \put(0.68,0.0){\color[rgb]{0,0,0}\makebox(0,0)[lb]{\smash{$\delta x$}}}%
  \end{picture}%
\endgroup%